\documentclass[twocolumn,amsmath,showkeys,amssymb,superscriptaddress,longbibliography,floatfix, pre]{revtex4-2}

\pdfoutput=1

\usepackage{bm}
\usepackage{yfonts}
\usepackage{graphicx,epsfig}
\usepackage{amsmath,amssymb,mathtools}
\usepackage[rgb]{xcolor}
\usepackage{hyperref}
\usepackage{braket}
\def\be{\begin{equation}}
\def\ee{\end{equation}}
\newcommand{\g}{\hat{g}}

\newcommand{\he}{\hat{e}}
\def\bc{\begin{center}}
\def\ec{\end{center}}
\def\bea{\begin{eqnarray}}
\def\eea{\end{eqnarray}}
\newcommand{\avg}[1]{\langle{#1}\rangle}
\newcommand{\Avg}[1]{\left\langle{#1}\right\rangle}

\begin{document}

\title{Emergent gravity from entropy}

\author{Ginestra Bianconi}
\email{ginestra.bianconi@gmail.com}
\affiliation{School of Mathematical Sciences, Queen Mary University of London, London, E1 4NS, United Kingdom}

\begin{abstract}
Emergence, the ability of effective field theories to capture a system's phenomenology at different scales, is a central theme in theoretical physics and statistical mechanics. In this work, Quantum Gravity from Entropy (QGfE)  is proposed as the high-energy completion of the Emergent Gravity from Entropy (EGfE) and thus, in the weak-field limit, General Relativity (GR) itself. In QGfE, the quantum gravity quest is addressed by introducing  the geometry-field (GF), a quantum field of dimension one that shadows the true metric. Specifically, the GF admits a non-zero fundamental state describing the geometry of the manifold by means of the GF-tetrad and GF-connection.  The curvature and the matter fields appear in the QGfE action only through their individual coupling with the fluctuations of the GF. 
Close to the fundamental state, the fluctuations of the GF  are massive with a mass given by  the Planck mass, while at the same time, the QGfE preserves a massless graviton.
The Emergent Gravity from Entropy (EGfE) action, which is expressed in terms of the Gravity from Entropy (GfE) action, is the effective theory of  QGfE, obtained for energy scales  of the GF much lower than the Planck scale. The proposed QGfE provides a quantum gravity framework to derive EGfE and thus, in its weak-field limit, classical GR, from an underlying  quantum field theory. The introduction of the GF emerges as a key element to achieve this goal, with its massive fluctuations providing insights for the interpretation of the dark sector.
\end{abstract}
\maketitle
\section{Introduction}

Since the discovery that black holes have an entropy~\cite{bekenstein1973black} and emit Hawking radiation~\cite{hawking1975particle,birrell1984quantum}, an important hypothesis in theoretical physics is that  statistical mechanics can offer relevant insights for reconciling quantum mechanics with gravity.
Different proposals of entropic gravity  have been formulated in the literature~\cite{jacobson1995thermodynamics,jacobson2016entanglement,dorau2025quantum,alonso2025spacetime,verlinde2011origin}, all based on the area law for the entropy of horizons. However, from a statistical mechanics point of view the area law should be interpreted as a thermodynamic relation, while reconciling quantum mechanics with gravity might require a statistical mechanics approach capturing the microscopic degrees of freedom of geometry.
The recently proposed Gravity from Entropy (GfE) theory~\cite{bianconi2025gravity,bianconi2026thermodynamics,bianconi2025quantum,bianconi2026lambda}  addresses the relation between statistical mechanics and gravity by proposing an alternative action for gravity, the GfE  action, which in the weak-field limit reduces to  General Relativity (GR).
The Lagrangian of the GfE action~\cite{bianconi2025gravity}  does not rely on the area law of horizons. Rather, it captures the information carried by the true metric that can be encoded in the metric induced by the matter fields and curvature, as expressed by the Geometric Quantum Relative Entropy (GQRE), related to the Araki entropy \cite{araki1975relative,witten2018aps} between these two metrics.  In this way, the GfE theory reveals that GR might have a statistical mechanics origin, and proposes the GfE action for capturing gravity at higher energy scales than GR. This framework has stimulated further research investigating the  applications of GfE to inflation~\cite{thattarampilly2025inflation},  black holes~\cite{bianconi2025quantum,pereira2026scalar}, the weak-field limit \cite{thattarampilly2026spherically,pereira2026vacuum}, and generalizations to other definitions of entropy~\cite{gohar2026gravity}.

Statistical mechanics is usually interpreted to be the theory of emergence that describes how macroscopic behavior emerges from the microscopic degrees of freedom of the systems, in the words of Phil Anderson {\em More is different}~\cite{anderson1972more}.
At the same time, however, there is a separate line of thought that postulates a fundamental information-theoretic character of Nature, beautifully summarized by John Wheeler's {\it It from bit}~\cite{wheeler2018information}. Thus, while interpreting  GfE, a central question emerges: {\em do we need to interpret GfE as a  fundamental theory of gravitation or as an emergent theory?}

This work originates from the belief that gravity should ultimately be described by a quantum field theory. This belief is shared by many but not by all ~\cite{oppenheim2023postquantum,angeli2025probing} in the community,  and will eventually require an experimental validation to be settled~\cite{bose2017spin,marletto2017gravitationally,bose2025massive}. 
Assuming that gravity is fundamentally quantum, the formulation of GfE within a first-quantization framework implies that it should necessarily be understood as an effective theory of a more fundamental quantum field theory. 

 In this work, a quantum gravity completion of GfE, called Quantum Gravity from Entropy (QGfE) is introduced. Consequently, here we revisit and reinterpret, in the GfE context, earlier ideas of induced gravity~\cite{Sakharov1967,donoghue1994general,visser2002sakharov}.
 We show that QGfE  admits the Emergent Gravity from Entropy (EGfE), which is formulated in terms of the GfE action, as its effective action. Since the classical EGfE reduces to General Relativity (GR) in the weak-field limit, QGfE  derives classical GR from a quantum field theory.


The QGfE  requires the introduction of a higher-order {\em geometry-field} (GF) $\Theta$ of dimension one. In the fundamental state, obtained by a gravitational symmetry breaking mechanism, the GF describes the true geometry of space-time with one of its components proportional to the true tetrad and another component proportional to the true Levi-Civita spin-connection of the  manifold. The  fluctuations of the GF, captured by the $\Phi$ field, couple with matter fields and curvature, which appear in the QGfE Lagrangian only through this coupling. Around the fundamental state of the QGfE action, the $\Phi$ field becomes massive with mass $M$ given by the Planck mass. The effective theory of QGfE leads to the EGfE action. Interestingly, the quantum fluctuations of the metric and the matter fields around their classical values are themselves driven by the variation of the EGfE action. 

The QGfE is here proposed as an approach to quantum gravity that addresses several of the field's main open questions~\cite{armas2021conversations}, placing it within the broader literature while offering several avenues for further research.

Firstly, the QGfE Lagrangian contains only dimensionless coupling constants. Thus, further exploring its renormalization group properties might provide further insights into the open problem of formulating a renormalizable theory of quantum gravity~\cite{wetterichberges2002non,rahmede2009investigating,falls2016further,falls2018asymptotic}.

Secondly, the QGfE action assumes that the true metric is torsion-free, but the GF is allowed to have a fluctuating GF-torsion, with the QGfE action containing a term that imposes as a soft constraint that the GF torsion is null. Consequently, in this way QGfE interprets the dichotomy present in modified theories of gravity that adopt either a torsion-free metric or affine formulations \cite{capozziello2011extended}.

Thirdly, close to its fundamental state, the field $\Phi$ captures the fluctuations of the GF, and explicitly represents the quantum fluctuations of the metric, which is a central physical concept in quantum gravity. Interestingly, the fluctuations of the GF are massive, with a Planck mass $M=M_P$ emerging through a gravitational symmetry breaking mechanism.  However, although the $\Phi$ field is massive, the QGfE continues to display a massless graviton. Thus, the QGfE action reveals a distinct way to address open quantum gravity questions that have been discussed in the context of proposed gravitational Higgs mechanisms~\cite{percacci1991higgs,maitiniyazi2025spacetime} and massive gravity~\cite{derham2011resummation,derham2014massive}. 

Finally, the QGfE displays EGfE as its effective classical theory,  which in the weak-field limit reduces to GR. Moreover, the  quantum fluctuations around this classical effective field theory are driven by the variation of the EGfE action itself.  Therefore, this work provides a path to quantum gravity which is alternative to other quantum gravity approaches~\cite{Oriti}, including string theories~\cite{polchinski2005string}, loop quantum gravity~\cite{Rovelli}, group field theory~\cite{freidel2005group}, and  asymptotic safety~\cite{rahmede2009investigating,falls2016further,falls2018asymptotic,Astrid}. 

Summing up, the QGfE defines a path to quantum gravity alternative to previous proposals, and  at the same time addressing open questions of the field. Specifically, the clear identification of the relation between QGfE and its effective field theory given by  EGfE, and the emergent role of EGfE itself to drive the quantum fluctuations around a classical background  might provide a fertile framework to test  this theory with astrophysical  results~\cite{berti2015testing,barausse2020prospects}. 
Moreover, the proposed theory also offers clues for the interpretation of the dark  sector \cite{karim2025desi,karim2025desi,carloni2025does,copeland2006dynamics,bahamonde2018dynamical,bertone2024dark,bertone2005particle} where the GF  contributes to the matter content of the Universe and has fluctuations that are responsible for the dark energy term of its GfE effective theory.

This paper is organized as follows: in Sec.~\ref{sec:nutshell} QGfE will be presented in a nutshell, together with the main results; in Sec. \ref{sec:Theta} the GF is introduced and its properties are discussed jointly with the properties of the true metric; in Sec.~\ref{sec:QGfE}  the QGfE action will be formulated and presented in detail; in Sec.~\ref{sec:quadratic}  the nature of the fundamental state of the QGfE and its quadratic expansion in the $\Phi$ field are derived; in Sec.~\ref{sec:neutral}  the leading term of the action that close to the fundamental state includes the matter fields is derived; in Sec.~\ref{sec:EGfE}  EGfE is derived as the effective field  theory of QGfE, first obtaining the classical EGfE and neglecting all quantum corrections due to the metric and the matter fields, and subsequently including the  contributions due to quantum fluctuations. Finally, in Sec. ~\ref{sec:conclusions}  the concluding remarks are provided. The paper also includes two Appendices providing more details on the derivation presented in the main paper.  

\begin{figure*}[!t]
    \includegraphics[width=1.98\columnwidth]{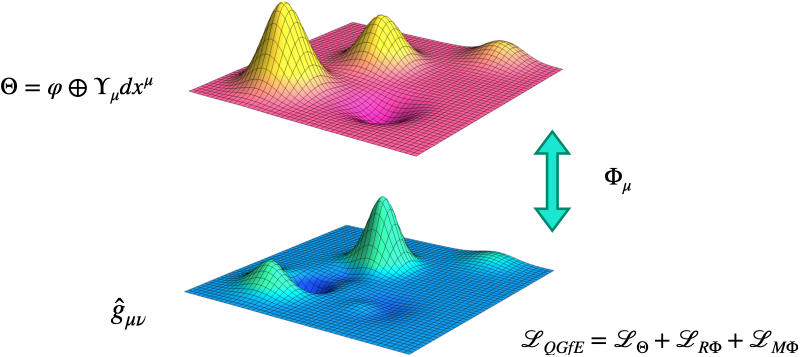}
    \caption{Illustration of the QGfE in a nutshell. The manifold $\mathcal{M}$ is associated with the true metric $\hat{g}_{\mu\nu}$ and the geometry-field (GF) $\Theta$. Through a symmetry breaking mechanism, driven by the free Lagrangian $\mathcal{L}_{\Theta}$, the GF acquires a non-zero value in the QGfE fundamental state, where  the GF-tetrad is proportional to the  true tetrad and the GF-connection reduces to  the Levi-Civita spin connection of the manifold $\mathcal{M}$. The fluctuations $\Phi$ of the GF are associated with the mass $M$, given by the Planck mass, i.e. $M=M_P$, and  couple with curvature and with matter fields  through the Lagrangians $\mathcal{L}_{R\Phi}$ and $\mathcal{L}_{M\Phi}$. The effective theory of QGfE is EGfE, which reduces to General Relativity (GR) in the weak-field limit.}
    \label{fig1}
    \end{figure*}
\section{The QGfE in a nutshell}
\label{sec:nutshell}
Quantum gravity addresses the quest for a quantum field theory that reconciles our understanding of gravity driven by the geometrical description of spacetime with the Standard Model. In order to address this challenge, QGfE introduces a higher-order complex-valued field, the {\em geometry-field} (GF), 
$\Theta$, which {\em shadows} the true metric.
By means of a gravitational symmetry breaking  mechanism for the emergence of geometry, the  free geometry-field $\Theta$  admits a non-zero fundamental state, in which one component of the GF, called GF-tetrad, reduces to the true tetrad multiplied by the Planck mass, and another component of the GF, called GF-connection, reduces to the Levi-Civita spin connection of the manifold. The fluctuations $\Phi$ of the GF  couple with matter and curvature, providing the means by which the matter fields and curvature interact with geometry (see Figure \ref{fig1} for an illustration of the theory).
The QGfE action $\mathcal{S}_{QGfE}$ is given by
\bea
\mathcal{S}_{QGfE}=\int d^4x \sqrt{-|\g|}\mathcal{L}_{{QGfE}} 
\eea
where the Lagrangian $\mathcal{L}_{{QGfE}}$ is given by
\bea
\mathcal{L}_{{QGfE}}=\mathcal{L}_{\Theta}+\mathcal{L}_{R\Phi}+\mathcal{L}_{M\Phi}.
\eea
The first term $\mathcal{L}_{\Theta}$ is the free Lagrangian for the GF $\Theta$; the other two terms $\mathcal{L}_{R\Phi}$ and $\mathcal{L}_{M\Phi}$ capture the coupling of the $\Phi$ field with the curvature and the matter fields, respectively. It  is to be emphasized that the whole QGfE Lagrangian $\mathcal{L}_{QGfE}$  does not contain any dimensional coupling constant and that both the curvature and the matter fields appear only in combination with the $\Phi$ field that acts as a kind of ``geometric glue" in addition to the true metric.

Close to the fundamental state, the fluctuations of the $\Theta$ field, captured by the $\Phi$ field, are massive with mass given by $M$, identified with the Planck mass, i.e. $M=M_P$.
The  effective theory of QGfE, obtained by integrating out the massive $\Phi$ fields, is driven by the Emergent Gravity from Entropy (EGfE)  action  $\mathcal{S}_{EGfE}$ given by
\bea
\mathcal{S}_{EGfE}=\mathcal{S}_{EGfE}^{(c)}+\mathcal{S}_{QC}
\eea
where the classical EGfE action $\mathcal{S}_{EGfE}^{(c)}$ neglects quantum fluctuations of the metric and the matter fields, while $\mathcal{S}_{QC}$ takes into account the perturbative quantum fluctuations of the metric and the  matter fields around the classical background.
The classical EGfE action $\mathcal{S}_{EGfE}^{(c)}$ is given by
\bea
\mathcal{S}^{(c)}_{EGfE}=\int d^4x\sqrt{-|g|}\mathcal{L}_{EGfE}^{(c)},
\eea
where the classical EGfE Lagrangian $\mathcal{L}^{(c)}_{EGfE}$  is given by the difference between the GfE Lagrangian $\mathcal{L}_{GfE}$ and the matter Lagrangian $\mathcal{L}_{M}$, i.e.
\bea
\mathcal{L}^{(c)}_{EGfE}=\mathcal{L}_{GfE}-\mathcal{L}_{M},
\eea
The  GfE  Lagrangian $\mathcal{L}_{GfE}$  \cite{bianconi2025gravity} is given by the GQRE between the true metric and the metric induced by the matter fields and curvature,
\bea
\mathcal{L}_{GfE}=-n_c\mbox{Tr}_F\ln (\tilde{\bf G}\tilde{g}^{-1}),
\eea
where $n_c=O(M_P^4)$ and the metric $\tilde{\bf G}$ induced by the matter fields and curvature is given by
\bea
\tilde{\bf G}=\tilde{g}+\alpha \tilde{\bf M}-\beta\tilde{\bm{\mathcal{R}}}.
\eea
Here, the constants $\alpha$ and $\beta$ are defined as 
\bea
\alpha=\frac{\alpha^{\prime}}{M^4},\quad \beta=\frac{\beta^{\prime}}{M^2},\label{eq:ab}\eea
where $\alpha^{\prime},\beta^{\prime}$ are dimensionless constants, while $\tilde{\bf M}$ and $\tilde{\bm{\mathcal{R}}}$ include the matter fields and  curvature contributions, respectively. Thus, GfE is here shown to emerge as a fundamental element of classical EGfE, i.e. the classical effective theory of QGfE. Note, however, that in this work  the explicit expression of the matter field content $\tilde{\bf M}$ is a modification of the one proposed in Ref. \cite{bianconi2025gravity}.
The emergent matter Lagrangian $\mathcal{L}_M$, also appearing in the classical EGfE action, is given by 
 \bea
\mathcal{L}_M=-\frac{n_c\alpha}{2}\mbox{Tr}_F\tilde{\bf M}\tilde{g}^{-1}.\eea
Thus, the classical EGfE can be, broadly speaking, interpreted as a free energy  with $\mathcal{L}_M$ playing the role of the energy and $\mathcal{L}_{GfE}$ playing the role of the entropy.
Finally, the term $\mathcal{S}_{QC}$ encodes the contributions coming from the quantum fluctuations of the metric and the matter fields.

\section{The true metric and the geometry field}
\label{sec:Theta}
\subsection{The true metric}
Throughout the paper we take units $c=\hbar=1$ and we indicate with $M_P$ the Planck mass.
We consider a $4$-dimensional torsion-free manifold $\mathcal{M}$ associated with a metric field $\hat{g}_{\mu\nu}$ with signature $(-,+,+,+)$.  
The manifold $\mathcal{M}$ has an internal Lorentz bundle ${V}$ with Minkowski metric $\eta_{ab}$, i.e. $(\mathbb{R}^{1,3},\eta_{ab})$.
The metric field $\hat{g}_{\mu\nu}$ of $\mathcal{M}$ can be expressed in terms of the tetrad field $\hat{e}_{\mu}^a$ and the Minkowski metric  $\eta_{ab}$ as
\bea
\hat{g}_{\mu\nu}=\hat{e}_{\mu}^a\hat{e}_{\nu}^b\eta_{ab}.
\eea
We should treat the tetrad field $\hat{e}_{\mu}^a$ and the Minkowski metric $\eta_{ab}$ as the fundamental fields that define $\hat{g}_{\mu\nu}$ and its associated Levi-Civita spin connection $\hat{\omega}_{\mu}^{ab}$. For the associated classical fields we drop the hat and adopt the standard notation $e_{\mu}^a,g_{\mu\nu}$. The curvature  is indicated as ${R}^{ab}_{\mu\nu}$, while the torsion is indicated as ${T}_{\mu\nu}^a$. 
\subsection{The geometry-field $\Theta$}
The geometry-field $\Theta$, of dimension $[\Theta]=1$, is the direct sum of a zero-form, complex-valued field  $\varphi\in \Omega_0(\mathcal{M})$ and a complex-valued, one-form $\Upsilon\in \tilde{\Omega}_1(\mathcal{M})$ with values in the internal Lorentz space.
Specifically, the geometry field $\Theta$  is given by  
\bea
\Theta=\varphi\oplus\Upsilon_{\mu}dx^{\mu}
\eea
with \bea
\Theta\in \Omega^0(\mathcal{M})\oplus \tilde{\Omega}^1(\mathcal{M},{V}).
\eea
The field $\Upsilon$  has a higher-order structure in the internal Lorentz space  and is expressed as 
\bea 
\Upsilon_{\mu}=\phi_{\mu}\oplus {\theta}_{\mu}^a dy_a\oplus\frac{1}{2}{\varsigma}_{\mu}^{ab} dy_a\wedge dy_b,
\eea
where $\theta_{\mu}^a$ is the GF-tetrad and $\varsigma_{\mu}^{ab}$ is the GF-connection. The reader is referred to Table $\ref{table1:Theta}$ for a schematic description of the role that the different components of the GF $\Theta$ play in the QGfE.
The $\Upsilon$ field lives in the space \bea\Upsilon\in \tilde{\Omega}^1=\Omega^1(\mathcal{M};\Lambda^0{V})\oplus\Omega^1(\mathcal{M};\Lambda^1{V})\oplus \Omega^1(\mathcal{M};\Lambda^2{V}).\nonumber\eea 
In $\tilde{\Omega}^1$ the metric is given by
\bea
\tilde{g}=\hat{g}_{\mu\nu}dx^{\mu}\otimes dx^{\nu}\otimes \tilde{\eta}\label{eq:gt}
\eea
with $\tilde{\eta}$ being the Minkowski metric between scalars, vectors and bivectors in $V$, i.e.
\bea
\tilde{\eta}&=&1\oplus \eta_{ab}dy^a\otimes dy^b\nonumber \\&&\oplus \frac{1}{2}\left(\eta_{ac}\eta_{bd}-\eta_{ad}\eta_{bc}\right)dy^{a}\wedge dy^{b}\otimes dy^{c}\wedge dy^{d}
\eea
Here and in the following, we will call $\theta_{\mu}^a$ the geometry-field tetrad (GF-tetrad) and $\varsigma_{\mu}^{ab}$ the geometry-field connection (GF-connection), since we will see that, in the fundamental state, the GF-tetrad is proportional to the true tetrad and the GF-connection reduces to the  Levi-Civita spin-connection of the manifold.

The $\varphi$ field is a {\em gravitational scalar field} that is responsible for the emergence of a non-zero GF-tetrad and a non-zero GF-connection in the fundamental state of QGfE from a gravitational symmetry breaking mechanism provided by its coupling with the one-form $\phi_{\mu}$. 

\begin{table*}[!ht!]
\caption{The components of the GF $\Theta$, their fundamental state and their role in QGfE.}
\begin{ruledtabular}
\begin{tabular}{lll}
field &    fundamental state & role\\
\hline
$\varphi\in\Omega_0(\mathcal{M})$,  & $|\varphi|=M/q$ where  $q$ is its charge & gravitational scalar, sets $M$\\
$\phi_\mu\in\Omega_1(\mathcal{M},\Lambda^0V)$ & 0 & gauge field of the phase of $\varphi$\\
$\theta_\mu{}^a\in\Omega_1(\mathcal{M},\Lambda^1V)$  & $M\he_\mu{}^a$ true tetrad $\he_\mu{}^a$, multiplied by $M$& GF tetrad \\
$\varsigma_\mu{}^{ab}\in\Omega_1(\mathcal{M},\Lambda^2V)$  & $\hat\omega_\mu{}^{ab}$  true Levi-Civita connection & GF-connection\\
\end{tabular}
\end{ruledtabular}
\label{table1:Theta}
\end{table*}

\subsection{The properties of the true metric and of the  geometry field}

The  torsion ${T}^a$ and the curvature ${R}^{ab}$ associated with the true metric are given by 
\bea
 {T}^a=d\he^a+\hat\omega^a{}_b\wedge\he^b,\qquad
  {R}^{ab}=d\hat\omega^{ab}+\hat\omega^a{}_c\wedge\hat\omega^{cb}.
\eea
The Levi-Civita connection is the unique connection for which the torsion ${T}^a$ vanishes, and constitutes the traditional way in which the differential geometry of the manifold is treated in General Relativity.
There are, however, affine gravity approaches ~\cite{capozziello2011extended} that relax the condition of having a torsion-free manifold, treating the connection and the tetrad as independent degrees of freedom of geometry to be determined by the equations of motion of modified gravity  actions.

In QGfE the true metric remains torsion-free, and is therefore associated with the Levi-Civita connection $\hat{\omega}_{\mu}^{ab}$. However,  the GF admits deviations from a torsion-free GF. Thus, the GF connection $\varsigma_{\mu}^{ab}$ generally deviates from the GF Levi-Civita spin connection $\hat{\varsigma}_{\mu}^{ab}$.

Since the GF tetrad $\theta_{\mu}^a$ is treated as an alternative tetrad, albeit with dimension $[\theta]=1$, we define its associated GF-torsion $\mathcal{T}^a$ of dimension $[\mathcal{T}]=2$ as 
 \bea
 \mathcal T^a=d\theta^a+\varsigma^a{}_b\wedge\theta^b
 \eea
 with components 
 \bea
  \mathcal T^a{}_{\mu\nu}=\partial_\mu\theta_\nu{}^a-\partial_\nu\theta_\mu{}^a+\varsigma_\mu{}^a{}_b\,\theta_\nu{}^b-\varsigma_\nu{}^a{}_b\,\theta_\mu{}^b ,
  \label{eq:TGF}
\eea
where $\varsigma_{\mu}^{ab}$ of dimension $[\varsigma_{\mu}^{ab}]=1$ is its associated GF connection. 
The GF contorsion $\zeta_{\mu}^{ab}$  is a tensor given by the difference between the GF connection and the Levi-Civita GF connection $\hat{\varsigma}_{\mu}^{ab}$, i.e.
\bea
\zeta_{\mu}^{ab}=\varsigma_{\mu}^{ab}-\hat{\varsigma}_{\mu}^{ab}.
\eea
In terms of the GF contorsion, we can write the GF torsion as 
\bea
\mathcal{T}^a=\zeta^{ab}\wedge \theta_{b},
\eea
From this expression, it is clear that, as expected,  for  $\zeta_{\mu}^{ab}=0$ the torsion vanishes, i.e. $\mathcal{T}^a_{\mu\nu}=0$. Thus,  a non-zero contorsion quantifies the deviation of the GF from a torsion-free GF.
For future reference, we also introduce $\mathcal{T}_{\lambda\mu\nu}$ where the first index is lowered by means of the tetrad $\hat{e}_{\lambda}^a$, i.e.
\bea 
\mathcal{T}_{\lambda\mu\nu}&=&\he_{\lambda}^bT^a{}_{\mu\nu}
\eta_{ab}.
\eea

\section{The QGfE action}\label{sec:QGfE}

\subsection{General properties of the QGfE action}
As already mentioned in Sec.~\ref{sec:nutshell}, the  QGfE has the action
\bea
\mathcal{S}_{QGfE}=\int d^4x \sqrt{-|\g|}\mathcal{L}_{{QGfE}} 
\eea
where the Lagrangian $\mathcal{L}_{{QGfE}}$ is the sum of the free GF Lagrangian $\mathcal{L}_{\Theta}$ and the  Lagrangians $\mathcal{L}_{R\Phi}$ and $\mathcal{L}_{M\Phi}$ expressing the coupling of the GF fluctuations $\Phi$ (see Table $\ref{table2:Phi}$ for a schematic illustration of its components) with matter and curvature, i.e.
\bea
\mathcal{L}_{{QGfE}}=\mathcal{L}_{\Theta}+\mathcal{L}_{R\Phi}+\mathcal{L}_{M\Phi}.
\label{eq:LQGfE}
\eea
The QGfE is diffeomorphism invariant and locally Lorentz invariant. Thus, despite the fact that, as we will see, the fluctuations of the GF are massive,  the fluctuations  of the true metric $\hat{g}$ in the weak-field limit are captured by a massless graviton.

\subsection{The free GF Lagrangian $\mathcal{L}_{\Theta}$}
\subsubsection{Decomposition of the free Lagrangian $\mathcal{L}_{\Theta}$}
The free Lagrangian $\mathcal{L}_{\Theta}$ depends only on the GF $\Theta$  and on the true metric.   
In this section, we will illustrate the role of  the free GF Lagrangian $\mathcal{L}_{\Theta}$ in describing the  GF as a  {\em shadow} of the true geometry. Indeed, anticipating already the results of  the next section, in the  fundamental state of this Lagrangian,   the GF-tetrad is proportional to the real tetrad and the GF-connection is given by the torsion-free Levi-Civita spin-connection of the manifold.
The free Lagrangian $\mathcal{L}_{\Theta}$ can be  decomposed into three contributions
\bea
\mathcal{L}_{\Theta}=\mathcal{L}_{\pi}+\mathcal{L}_{\kappa}+\mathcal{L}_{\mathcal{T}}.
\eea
In the next paragraphs we will illustrate the physics behind each of these terms separately.
\subsubsection{The Lagrangian term $\mathcal{L}_{\pi}$}
The first term $\mathcal{L}_{\pi}$ of the free QGfE Lagrangian $\mathcal{L}_{\Theta}$ expresses the coupling of the 
 fields $\varphi$ and  $\phi_{\mu}$  by means of a  gravitational Higgs mechanism  responsible for a non-zero fundamental state for the GF and is given by 
 \bea
 \mathcal{L}_{\pi}=-\frac{1}{2}\,\bar f_{\mu\nu}f^{\mu\nu}-\big|D_\phi\varphi\big|^2-\lambda\Big(q^2|\varphi|^2-M^2\Big)^2,
 \eea
 where $D_{\phi}$ and $f_{\mu\nu}$ are defined as  
 \bea
D_\phi\varphi&=&(\partial_\mu-i q\phi_\mu)\varphi,\nonumber \\
   f_{\mu\nu}&=&\partial_\mu\phi_\nu-\partial_\nu\phi_\mu. 
  \label{eq:CPJ}
\eea
Accordingly,  $q$ can be interpreted as the charge  of the gravitational scalar $\varphi$.
We note that  $\mathcal{L}_{\pi}$ is invariant under the gauge transformations
\bea
\varphi\to \varphi e^{i\alpha}\quad \phi_{\mu}\to \phi_{\mu}+\frac{1}{q}\partial_{\mu}\alpha.
\eea
where $\alpha(x)\in \mathbb{R}$. Note that, since $\phi_{\mu}$ is complex-valued, only the real part of $\phi_{\mu}$ plays the role of a gauge field in this case.
We introduce a gauge-invariant complex-valued Proca field $\pi_{\mu}$ given by
\bea
\pi_\mu\equiv\phi_\mu+\frac{i}{q}\,\partial_\mu\ln\varphi.
\eea
which allows us to  express the Lagrangian $\mathcal{L}_{\pi}$ as
\bea
  \mathcal L_\pi=-\frac{1}{2}|f^{(\pi)}_{\mu\nu}|^2-q^2|\varphi|^2\,\bar \pi_\mu \pi^\mu-\lambda\Big(q^2|\varphi|^2-{M^2}\Big)^2,
\eea
where 
\bea
f^{(\pi)}_{\mu\nu}=\partial_{\mu}\pi_{\nu}-\partial_{\nu}\pi_{\mu}.
\eea
Thus, close to the  fundamental state, where  $|\varphi|^2=M^2/q^2$,  the Proca field $\pi_{\mu}$ acquires mass $M$.
Specifically, for $|\varphi|^2\simeq M^2/q^2$, the Lagrangian $\mathcal{L}_\pi$ can be written in terms of the Ricci tensor $R^{\mu\nu}$ by taking into account the  Weitzenb\"ock term, as 
\bea
\mathcal{L}_\pi=\bar{\pi}_{\mu}(\square \hat{g}^{\mu\nu}-\hat{R}^{\mu\nu}-M^2\hat{g}^{\mu\nu})\pi_{\nu}.
\eea
In this limit, the complex Proca field $\pi_{\mu}$ has six healthy degrees of freedom, with the time components fixed by the constraint $\nabla^{\mu}\pi_{\mu}=0$. Thus, in this way, the Proca field absorbs  the $4$ degrees of freedom of the massless $\phi_{\mu}$ field and the $2$ degrees of freedom of the complex scalar field $\varphi$.
\subsubsection{The Lagrangian term $\mathcal{L}_{\kappa}$}
The second term $\mathcal{L}_{\kappa}$ of the free QGfE Lagrangian $\mathcal{L}_{\Theta}$ imposes as a soft constraint that the GF tetrad is proportional to the true tetrad. Specifically, 
\bea
\mathcal{L}_{\kappa}=-{q^2}|\varphi|^2\bar\kappa_{\mu}^a\kappa^{\mu}_a,
\eea
where  $\kappa_{\mu}^a$ is defined as 
\bea
\kappa_{\mu}^a&=&{\theta}^{\mu}_a-q|\varphi|\he^{\mu}_a.
\eea
Therefore, $\mathcal{L}_\kappa$ imposes as a soft constraint
 \bea
 {\theta}^{\mu}_a=q|\varphi|\he^{\mu}_a.
 \eea
This implies that in the fundamental state $\theta^{\mu}_a=M\he_{\mu}^a$.
\subsubsection{The Lagrangian term $\mathcal{L}_{\mathcal{T}}$}
Finally, the third term $\mathcal{L}_{\mathcal{T}}$ of the free QGfE Lagrangian $\mathcal{L}_{\Theta}$ vanishes if the GF torsion is null, and is given by 
\bea
\mathcal{L}_{\mathcal{T}}=-\frac{1}{4}\bar{\mathcal{K}}_{\mu\nu\rho}\mathcal{K}^{\mu\nu\rho}
\eea
where $\mathcal{K}_{\mu\nu\rho}$ is the Koszul combination and is given by 
\bea
\mathcal{K}_{\mu\nu\rho}&=&\mathcal{T}_{\mu\nu\rho}+\mathcal{T}_{\nu\mu\rho}+\mathcal{T}_{\rho\nu\mu}.
\eea
The Koszul combination $\mathcal{K}_{\mu\nu\rho}$ can be expressed as 
\bea
\mathcal{K}_{\mu\nu\rho}=2\zeta_\mu{}^{ab}\Sigma_{\nu\rho}^{cd}\eta_{ac}\eta_{bd} +2\big(S_{\nu\mu\rho}-S_{\rho\mu\nu}\big),
\eea
where 
\bea
\Sigma_{\mu\nu}{}^{ab}=\left(\he_{\mu}{}^a\theta_{\nu}{}^b-\he_{\nu}{}^a\theta_{\mu}{}^b\right),\nonumber \\
S_{\mu\lambda\nu}=\frac{1}{2}\zeta_\mu{}^{ab}\left(\he_{\lambda a}\kappa_{\nu b}+\he_{\nu a}\kappa_{\lambda b}\right)
\eea
It follows that $\mathcal{L}_{\mathcal{T}}$ can be simultaneously interpreted as a kinetic term for the GF tetrad and as a soft constraint for the auxiliary GF contorsion. Indeed, the torsion Lagrangian $\mathcal{L}_{\mathcal{T}}$ is quadratic in the contorsion $\zeta_{\mu}^{ab}$, and is non-degenerate as long as $\theta_{\mu}^a$ is invertible. Thus, since $\zeta_{\mu}^{ab}$ is auxiliary and does not have a kinetic term,  in the fundamental state we obtain $\zeta_{\mu}^{ab}=0$ and thus a vanishing GF torsion $\mathcal{T}^a_{\mu\nu}=0$.

We also note that the Lagrangian $\mathcal{L}_{\mathcal{T}}$ can  equivalently be expressed in terms of the GF-torsion invariants as 
\bea
\mathcal{L}_{\mathcal{T}}=-\frac{1}{4}\Big[3\,\bar{\mathcal{T}}_{\lambda\mu\nu}\mathcal{T}^{\lambda\mu\nu}+2\,\bar{\mathcal{T}}_{\lambda\mu\nu}\mathcal{T}^{\mu\lambda\nu}\Big].
\eea

\subsubsection{Comprehensive expression of the free Lagrangian}
Summing up all the terms, we can  express the free GF Lagrangian $\mathcal{L}_{\Theta}$ as 
\bea
\mathcal L_{\Theta}&=&-\frac{1}{2}\,\bar f_{\mu\nu}f^{\mu\nu}-\big|D_\phi\varphi\big|^2-\lambda\Big(q^2|\varphi|^2-M^2\Big)^2\nonumber \\&&-{q^2}|\varphi|^2\bar\kappa_{\mu}^a\kappa^{\mu}_a-\frac{1}{4}\bar{\mathcal{K}}_{\mu\nu\rho}\mathcal{K}^{\mu\nu\rho}.\eea
\begin{table*}[!htb!]
\caption{The components of the field $\Phi$, which couples the GF to matter and curvature, their definition in terms of the GF and their role in QGfE.}
\begin{ruledtabular}
\begin{tabular}{lll}
field &    definition & role\\
\hline
$\pi_\mu\in\Omega_1(\mathcal{M},\Lambda^0V)$ & $\pi_\mu\equiv\phi_\mu+\frac{i}{q}\,\partial_\mu\ln\varphi$
 & gauge-invariant field \\
$\kappa_\mu{}^a\in\Omega_1(\mathcal{M},\Lambda^1V)$  & $\kappa_{\mu}^a=\theta_{\mu}^a-q|\varphi|\he_{\mu}^a$ & deviation of GF tetrad from its fundamental state \\
$\zeta_\mu{}^{ab}\in\Omega_1(\mathcal{M},\Lambda^2V)$  & $\zeta_{\mu}^{ab}=\varsigma_{\mu}^{ab}-\hat{\varsigma}_{\mu}^{ab}$ & GF-contorsion\\
\end{tabular}
\end{ruledtabular}
\label{table2:Phi}
\end{table*}
\subsection{The $\Phi$ field}
The coupling between  matter and curvature is mediated by the complex-valued field $\Phi\in \tilde{\Omega}_1$ which describes the fluctuations of the GF and is given by
\bea
\Phi=\Phi_{\mu}dx^{\mu}
\eea
where $\Phi_{\mu}$ is valued in the internal Lorentz space and is defined as 
\bea
\Phi_{\mu}=\pi_{\mu}\oplus \kappa_{\mu}^ady_a\oplus\frac{1}{2} \zeta_{\mu}^{ab}dy_a\wedge dy_b.
\eea
Here and in the following we assign a {\em gravitational charge} $q_G=+1$ to $\Phi$ and $q_G=-1$ to $\bar\Phi$.
For a schematic illustration of the different roles of the components of the $\Phi$ field, within the QGfE theory,  we refer the reader to Table \ref{table2:Phi}.
In the QGfE the $\Phi$ field provides the coupling with  curvature and matter as determined by the Lagrangians $\mathcal{L}_{R\Phi}$ and $\mathcal{L}_{M\Phi}$, both quadratic in the $\Phi$ field.
\subsection{Coupling with curvature}
In QGfE the  GF is coupled to the true metric in two ways. On the one hand, the term $\mathcal{L}_{\kappa}$ in the Lagrangian $\mathcal{L}_{\Theta}$ depends quadratically on $\kappa_\mu^a$ that quantifies the difference between the GF tetrad and the true tetrad multiplied by $q|\varphi|$. On the other hand, the  fluctuations of the GF, captured by the $\Phi$ field, are coupled with the curvature tensor $R_{\mu\nu\rho\sigma}$ through the Lagrangian $\mathcal{L}_{R\Phi}$, present in the $QGfE$ Lagrangian (Eq.(\ref{eq:LQGfE})).

The Lagrangian $\mathcal{L}_{R\Phi}$ is given by the following coupling of the $\Phi$ field with the Ricci scalar $R$, the Ricci tensor $R_{\mu\nu}$ and the Riemann tensor $R_{\mu\nu\rho\sigma}$,  i.e.
\bea
\mathcal{L}_{R\Phi}=\beta^{\prime}\bar{\Phi}^{\mu}\hat{\bm{\mathcal{R}}}\Phi_{\mu},
\label{eq:LRPhi}
\eea
where 
\bea
\hat{\bm{\mathcal{R}}}&=&a_0R\oplus a_1 R_{ab}dy^a\otimes dy^b\nonumber \\
&&\oplus R_{ab}{}{}^{cd} dy^a\wedge dy^b\otimes dy_c\wedge dy_d.\label{eq:hR}
\eea
Here $\beta^{\prime}>0,a_0\geq 0,a_1\geq 0$ are  dimensionless constants of QGfE.
Thus,  the $\mathcal{L}_{R\Phi}$ Lagrangian can be explicitly written in terms of the components of the $\Phi$ field as 
\bea
\hspace{-3mm}\mathcal{L}_{R\Phi}=\beta^{\prime}\left[a_0\bar{\pi}_{\mu}R\pi^{\mu}+a_1\bar\kappa_{\mu}^aR_{ab}\kappa^{\mu}_b+\bar{\zeta}^{ab}_{\mu}R_{ab}{}{}^{cd}\zeta_{cd}^{\mu}\right].
\eea

\subsection{Coupling with matter}
Although ideally we would desire to include the whole Standard Model in the QGfE, in this paper we start by considering the   Lagrangian $\mathcal{L}_{M\Phi}$ determining  the coupling with  a scalar bosonic field $\chi$, and an Abelian field $A$. The extension of the proposed framework to fermionic matter is left for future work.

The coupling with the considered bosonic matter fields is driven by the Lagrangian $\mathcal{L}_{M\Phi}$ given by
\bea
\mathcal{L}_{M\Phi}&=&{\sqrt{\alpha'}}\,\pi^{\nu}\left(i \frac{1}{\sqrt{2}}m_\chi\pi_\nu|\chi|+\kappa^{a}_{\nu}\hat{e}^{\rho}_a\bar{\nabla}_{\rho}\chi+\frac{i}{2}\zeta^{ab}_{\nu}F_{ab}\right)\nonumber \\
&&\hspace{-12mm}-{\sqrt{\alpha'}}\,\bar\pi^{\nu}\left(-i \frac{1}{\sqrt{2}}m_\chi\bar\pi_\nu|{\chi}|+\bar\kappa^{a}_{\nu}\hat{e}_a^{\rho}\bar{\nabla}_{\rho}\bar\chi-\frac{i}{2}\bar\zeta^{ab}_{\nu}F_{ab}\right),\label{eq:LB}
\eea
where $m_{\chi}$ indicates the mass of the scalar and the covariant derivative $\bar{\nabla}_{\mu}\chi$ depends on the charge $q_{\chi}$ of the scalar,  
\bea
\bar{\nabla}_{\mu}\chi=\partial_{\mu}\chi-iq_{\chi}A_{\mu}\chi,\eea
while $F_{ab}$ is defined as 
\bea
F_{ab}=\left(\partial_{\mu}A_{\nu}-\partial_{\nu}A_{\mu}\right)\hat{e}_{[a}^{\mu}\hat{e}_{b]}^{\nu}.
\eea
For future reference, we define the vectors  ${\bf \Phi}_{\mu}$  and ${\bf J}$  given by 
\bea
{\bf \Phi}_{\mu}&=&\left(\pi_{\mu},\kappa_{\mu}^{a},\zeta_{\mu}^{ab}\right)\nonumber \\
{\bf J}&=&\left(\frac{i}{\sqrt{2}}\,m_\chi|\chi|,\hat{e}^{\rho}_a \bar\nabla_\rho\chi,\frac{i}{2}F_{ab}\right),
\label{eq:J}
\eea
of generic components $\Phi_{\mu}^N$ and $J_N$. 
In terms of these components, the   Lagrangian $\mathcal{L}_{M\Phi}$ can be alternatively expressed as  
\bea
\mathcal{L}_{M\Phi}=\sqrt{\alpha'}\,(\pi^\nu\,\Phi^N_\nu J_N-\bar\pi^\nu\,\bar{\Phi}^N_\nu \bar{J}_N).
\eea
From the expression of the Lagrangian $\mathcal{L}_{M\Phi}$, it is apparent that the bosonic fields  couple linearly with quadratic terms in the $\Phi$ fields, and that the Lagrangian $\mathcal{L}_{M\Phi}$ is chosen to be anti-Hermitian.  
We note, however, that, since we  assign a gravitational charge $q_G=+1$ to $\Phi$ and a charge $q_G=-1$ to $\bar{\Phi}$, the matter Lagrangian includes a term, $\sqrt{\alpha'}\pi^\nu\,\Phi^N_\nu J_N$, with charge $q_G=+2$ and a  term, $-\sqrt{\alpha'}\bar{\pi}^\nu\bar{\Phi}^N_\nu \bar{J}_N$, of charge $q_G=-2$. Since the free GF action conserves this gravitational charge, only charge-neutral couplings contribute to the QGfE effective action. Consequently, the leading neutrally charged matter contribution to the quadratic action in the $\Phi$ field is Hermitian and quadratic in the matter fields $\chi$ and $A$.

\section{The  fundamental state of QGfE and the QGfE  expansion}
\label{sec:quadratic}
\subsection{The fundamental state of QGfE}
The fundamental state of QGfE is achieved for $\Phi=0$ where the components of the GF $\Theta$ take the values 
  \bea
  |\varphi|=\frac{M}{q}, \qquad \phi_\mu=0,\nonumber \\
  \theta_{\mu}^a=M \he_{\mu}^a,\qquad \varsigma_{\mu}^{ab}=\hat{\omega}_{\mu}^{ab}.
\eea
for any given metric $\hat{g}_{\mu\nu}$.
In this  state, the scalar field of geometry $\varphi$ takes a non-zero absolute value $M/q$, breaking the symmetry of the problem, and consequently  the GF-tetrad $\theta_{\mu}^a$ reduces to  the Planck mass $M$ multiplied by the true tetrad $\hat{e}_{\mu}^a$, and the GF-connection $\varsigma_{\mu}^{ab}$ reduces to the  true Levi-Civita spin-connection $\hat{\omega}_{\mu}^{ab}$ of the manifold $\mathcal{M}$. In this case, the $\Phi$ field vanishes, i.e. $\Phi=0$. Thus,  the coupling between matter and curvature is no longer mediated by the $\Phi$ field and the metric is not determined. Only the fluctuations of the GF captured by a non-zero $\Phi$ field will allow  the metric to be determined.

\subsection{The quadratic expansion of the QGfE Lagrangian}

Developing the free GF Lagrangian $\mathcal{L}_{\Theta}$ close to the fundamental state, i.e. for a small $\Phi_{\mu}$ field, we get that, up to quadratic terms in the $\Phi$ field, the free GF Lagrangian is well approximated by the Lagrangian $\mathcal{L}_{\Phi}$ which is quadratic in the $\Phi$ field, i.e. $\mathcal L_{\Theta}\simeq \mathcal{L}_{\Phi}$ with
\bea
\mathcal{L}_{\Phi}=-\frac{1}{2}|f^{(\pi)}_{\mu\nu}|^2-M^2(\bar\pi_{\mu}\pi^{\mu}+\bar\kappa_{\mu}^a\kappa^{\mu}_a+\bar\zeta_{\mu}^{ab}\zeta^{\mu}_{ab}). 
\label{eq:Lkappa}
\eea
In this limit, the $\pi_{\mu}$ field is a propagating massive Proca field, while the $\kappa_{\mu}^a$ and the $\zeta_{\mu}^{ab}$ are massive but auxiliary and not propagating. All components of the $\Phi$ field have the same mass $M=M_P$.
Therefore, taking $\nabla^{\mu}\pi_{\mu}=0$, $\mathcal L_{\Phi}$ demonstrates that all the relevant degrees of freedom (dof)  of the $\Phi$ field (see Table $\ref{table3:dof}$ for a schematic overview) are massive. Indeed, close to the fundamental state,  $\mathcal{L}_{\Phi}$ can  be written in terms of the $\Phi$ field as 
\bea
\mathcal{L}_{\Phi}=\bar{\Phi}_{\mu}(\tilde{\square}\hat{g}^{\mu\nu}-\mathbb{R}^{\mu\nu}-M^2\tilde{g}^{\mu\nu})\Phi_{\nu},
\label{eq:LPhi}
\eea
where we have separated the Weitzenb\"ock term for the Proca Lagrangian, and $\tilde{\square}$ and $\mathbb{R}$ are given by
\bea
\tilde{\square}&=&\square \oplus\  0_{ab} \ dy^a\otimes dy^b\nonumber \\&&\oplus \ 0_{abcd}\ dy^a\wedge dy^b\otimes dy^c\wedge dy^d,\nonumber\\
\mathbb{R}^{\mu\nu}&=&\hat{R}^{\mu\nu}\oplus\  0^{\mu\nu}_{ab}dy^a\otimes dy^b\nonumber \\
&&\oplus\  0^{\mu\nu}_{abcd}dy^a\wedge dy^b\otimes dy^c\wedge dy^d.
\eea
The propagator associated with the quadratic fluctuations of $\Phi$ is therefore given by 
\bea
\Delta_{\mu\nu}(x-y)= i\Big[\tilde{\square}\hat{g}-\mathbb{R}-M^2\tilde{g}\Big]^{-1}_{\mu\nu}\frac{\delta^4(x-y)}{\sqrt{-|\hat g|}}.
\eea
For energy scales smaller than $\epsilon M$ with $\epsilon\ll 1$ we can approximate $\Delta_{\mu\nu}(x-y)$ by the uniform propagator given by its leading term
\bea
\Delta_{\mu\nu}(x-y)=-i\frac{1}{M^2}\tilde{g}_{\mu\nu}\frac{\delta^4(x-y)}{\sqrt{-|\hat g|}}.\label{eq:Dprop}
\eea
where, for brevity, we denote its components by \bea\Delta_{\mu\nu}^{N N^{\prime}}(x-y)=\Avg{0|T(\bar{\Phi}_{\mu}^N(x)\Phi_{\nu}^{N^{\prime}}(y))|0},\eea
and with $\Delta(x-y)$ simply 
\bea
\Delta(x-y)=-i\frac{1}{M^2}\frac{\delta^4(x-y)}{\sqrt{-|\hat g|}}.
\eea
\subsection{Quadratic expression of $\mathcal{L}_{QGfE}$ close to the fundamental state}
We conclude this section by observing that, close to the fundamental state,  the QGfE action $\mathcal{S}_{QGfE}$ can be approximated by $\mathcal{S}_{QGfE}^{(2)}$, associated with a Lagrangian that is quadratic in the  $\Phi$ field. Specifically, $\mathcal{S}_{QGfE}^{(2)}$ is  given by 
\bea
\mathcal{S}_{QGfE}^{(2)}=\int d^4x\sqrt{-|\hat{g}|}{\mathcal{L}}_{QGfE}^{(2)},\nonumber
\label{eq:S2}
\eea
with 
 \bea
\mathcal{L}_{QGfE}^{(2)}=\hat{\mathcal{L}}_{R\Phi}+\mathcal{L}_{M\Phi}.
\eea
Here $\mathcal{L}_{M\Phi}$ is given by Eq.(\ref{eq:LB}) while $\hat{\mathcal{L}}_{R\Phi}$ is given by 
\bea\hat{\mathcal{L}}_{R\Phi}=\mathcal{L}_{\Phi}+\mathcal{L}_{R\Phi}\eea
Consequently, using Eq.(\ref{eq:LPhi}) and Eq.(\ref{eq:LRPhi}), we can express $\hat{\mathcal{L}}_{R\Phi}$ as \bea
\hat{\mathcal{L}}_{R\Phi}=\bar{\Phi}^{\mu}(\tilde{\square}\hat{g}_{\mu\nu}-M^2\tilde{g}_{\mu\nu}+\beta^{\prime}\tilde{\bm{\mathcal{R}}}_{\mu\nu})\Phi^{\nu},\eea
with $\tilde{\bm{\mathcal{R}}}$ including both the curvature kernel $\hat{\bm{\mathcal{R}}}$ defined in Eq.(\ref{eq:hR}) and the Weitzenb\"ock term $\mathbb{R}$, i.e.
\bea
\tilde{\bm{\mathcal{R}}}_{\mu\nu}=\hat{\bm{\mathcal{R}}}\hat{g}_{\mu\nu}-\frac{1}{\beta^{\prime}}\mathbb{R}_{\mu\nu}.
\eea
The Lagrangian $\hat{\mathcal{L}}_{R\Phi}$ is both quadratic in the $\Phi$ field and neutral in the gravitational charge, as it contains exactly one factor of $\Phi$ (contributing +1 to the gravitational charge $q_G$) and one factor $\bar{\Phi}$ (contributing -1 to the gravitational charge $q_G$).
We note, however,  that the Lagrangian $\mathcal{L}_{M\Phi}$, albeit quadratic in the $\Phi$ field, is not neutral in the gravitational charge, as it contains terms of gravitational charge $q_G=+2,-2$. Thus, in the next section our aim will be to express the leading neutrally charged term, which depends on the matter fields and contributes to the QGfE partition function in the limit of small $\Phi$ field.

\begin{table}[t]
\caption{Degrees of freedom (dof) of  GF  close to the fundamental state}
\begin{ruledtabular}
\begin{tabular}{lcl}
field & dof & nature\\
\hline
$\varphi$ & $2$ &  complex scalar field\\
$\phi_\mu$ & $4$& massless complex vector field\\
$\pi_{\mu}$  & $6$ & complex Proca field, mass $M$,\\
& & (absorbing  $\varphi$ and $\phi_{\mu}$ dof)\\
$\kappa_\mu^a$ & $32$ & auxiliary, mass $M$\\
$\zeta_\mu^{ab}$ & $48$ & auxiliary, mass $M$\\
\end{tabular}
\end{ruledtabular}
\label{table3:dof}
\end{table}

\section{The leading neutrally charged contribution of the matter fields}
\label{sec:neutral}
In the previous section, we have derived the action $\mathcal{S}_{QGfE}^{(2)}$ from the QGfE action while retaining only quadratic terms in the $\Phi$ field in the expansion around the fundamental state in which the $\Phi$ field vanishes, i.e. $\Phi=0$. However, the action $\mathcal{S}_{QGfE}^{(2)}$ given by Eq.(\ref{eq:S2})  contains the term $\mathcal{S}_{M\Phi}$ that is not neutrally charged. Since only the neutrally charged free propagators of the field $\Phi$ are non-zero, in this section  we  use the Hartree approximation to obtain the leading neutrally charged quadratic contribution coming from the matter fields.

By indicating all  matter fields  as $\hat{\Psi}$, we  define the partition function $Z_{\Phi}[\hat\Psi,\tilde{\mathcal{R}},\g]$ which integrates the  action $\mathcal{S}_{QGfE}^{(2)}$ over the heavy field $\Phi$, i.e.
\bea
Z_{\Phi}[\hat\Psi,\tilde{\mathcal{R}},\g]=\int \mathcal{D}\bar\Phi\int \mathcal{D}\Phi e^{{i}\mathcal{S}_{QGfE}^{(2)}}.
\label{Z}
\eea

Since only the terms with neutral gravitational charge survive, we expand $e^{i\mathcal{S}_{M\Phi}}$ for a small $\Phi$ field. Noting that the first non-zero contribution in this expansion is quadratic in $\mathcal{S}_{M\Phi}$, we express the partition function $Z_{\Phi}[\hat{\Psi},\tilde{\mathcal{R}},\hat{g}]$ as 
\bea
Z_{\Phi}[\hat\Psi,\tilde{\mathcal{R}},\g]=\int \mathcal{D}\bar\Phi\int \mathcal{D}\Phi e^{{i}(\hat{\mathcal{S}}_{R\Phi}+{\mathcal{S}}^{\prime}_{M\Phi})}
\label{eq:Zprime}
\eea
where ${\mathcal{S}}_{M\Phi}^{\prime}$ is a bilocal effective action
 that can be expressed retaining only the term with zero gravitational  charge. 
The effective  bilocal action ${\mathcal{S}}_{M\Phi}^{\prime}$  is given by
\bea
{\mathcal{S}}_{M\Phi}^{\prime}=\int d^4x \sqrt{-|\g|}\int d^4y \sqrt{-|\g|}\mathcal{F}_B(x,y)
\eea
with $\mathcal{F}_B(x,y)$ given by
\bea
\mathcal{F}_B(x,y)=-{i}{\alpha^{\prime}}\left[\bar{\pi}_{\mu}(x)\bar{\Phi}^{\mu}_{N^{\prime}}(x) \bar{J}^{N'}(x)\pi_{\nu}(y)\Phi^{\nu}_N(y){J}^N(y) \right].\nonumber \eea

By considering the Hartree approximation, applied to the products  of type $\bar\Phi\bar\Phi\Phi\Phi$, present in $\mathcal{S}_{M\Phi}^{\prime}$, and using the large-$M$ expansion of the propagators of the $\Phi$ field, given by Eq.(\ref{eq:Dprop}), we reduce the bilocal action ${\mathcal{S}}_{M\Phi}^{\prime}$ (see Appendix $\ref{ApA}$ for details) to a local action $\hat{\mathcal{S}}_{M\Phi}$, i.e.
\bea
{\mathcal{S}}_{M\Phi}^{\prime}\simeq \hat{\mathcal{S}}_{M\Phi}=\int d^4x\sqrt{-|\g|}[\hat{\mathcal{L}}_{M\Phi}(x)-\hat{\mathcal{L}}_M(x)]\label{eq:SphS}
\eea where $\hat{\mathcal{L}}_{M\Phi}$ is  gravitationally neutral and quadratic in the $\Phi$ field, while, as a consequence of the Hartree approximation, the Lagrangian $\hat{\mathcal{L}}_M$, independent of the $\Phi$ field and  depending on the matter fields, also contributes to the action ${\mathcal{S}}_{M\Phi}^{\prime}$. 
Let us discuss first the Lagrangian $\hat{\mathcal{L}}_{M\Phi}$ which  is given by 
\bea
\hat{\mathcal{L}}_{M\Phi}&=&-\alpha M^2 \bar\Phi\tilde{\bf M}\Phi,
\eea
where we recall that $\alpha=\alpha^{\prime}/M^4$, and the flattened metric is defined as in Ref.\cite{bianconi2025gravity}. 
Here $\tilde{\bf M}$, which includes  the dependence on all the matter fields,  is given by 
\begin{widetext}
\bea
\tilde{\bf M}&=&\begin{pmatrix}
2m_\chi^2|\chi|^2+|\bar\nabla\chi|^2+\frac{1}{4}F_{\rho\sigma}F^{\rho\sigma} & -i{\sqrt{2}}m_\chi\bar\chi\he^\sigma_c(\bar\nabla_\sigma\chi) & \frac{1}{\sqrt{2}}m_\chi\bar\chi F_{cd}\\
i{\sqrt{2}} m_\chi \he^\rho_a(\bar\nabla_\rho\bar\chi)\chi & \he^\rho_a\bar\nabla_\rho\bar\chi\;\he^\sigma_c\bar\nabla_\sigma\chi & \frac i2\,\he^\rho_a\bar\nabla_\rho\bar\chi F_{cd}\\
\frac{1}{\sqrt{2}}\,m_\chi F_{ab}\chi & -\frac{i}{2}F_{ab}\;\he^\sigma_c\bar\nabla_\sigma\chi & \frac{1}{4}F_{ab}F_{cd}
\end{pmatrix},\eea
\end{widetext}

A relevant role in exploring the effective theory of QGfE is played by the trace of $\tilde{\bf M}$ which is proportional to the standard  matter Lagrangian, i.e.
\bea
\mbox{Tr}_F\tilde{\bf M}\tilde{g}^{-1}&=&8\left[ m_\chi^2|\chi|^2+|\bar\nabla\chi|^2+\frac{1}{4}F_{\mu\nu}F^{\mu\nu}\right].
\label{eq:TrM}
\eea
This term determines the Lagrangian $\hat{\mathcal{L}}_M$ which is the Hartree constant term (i.e. independent of the $\Phi$ field). Specifically, we evaluate $\hat{\mathcal{L}}_M$  for large mass $M$ using the propagator of the $\Phi$ field given by Eq.(\ref{eq:Dprop}) and integrating over the momentum of the $\Phi$ fields with upper cutoff $\epsilon M$, getting
\bea
\hat{\mathcal{L}}_{M}&=&i\frac{\alpha n_c}{2}\mbox{Tr}_F\tilde{\bf M}\tilde{g}^{-1}.
\eea
Here and in the following $n_c$  indicates the number of field modes per unit proper volume, integrated with the cutoff $\epsilon M$ consistently with the previous approximations made. In particular  (see Appendix \ref{ApB} for the derivation) $n_c$ is given by
\bea
n_c=\frac{\epsilon^4M^4}{6\pi^3}.
\eea

Substituting Eq.(\ref{eq:SphS}) into Eq.(\ref{eq:Zprime}), it follows  that the partition function $Z_{\Phi}[\hat{\Psi},\tilde{\mathcal{R}},\hat{g}]$, when retaining only the leading neutrally charged terms, can be written as 
\bea
Z_{\Phi}[\hat{\Psi},\tilde{\mathcal{R}},\hat{g}]=\int \mathcal{D}\bar\Phi\int \mathcal{D}\Phi e^{{i}\hat{\mathcal{S}}_{QGfE}},
\label{eq:Zh}
\eea
with action $\hat{\mathcal{S}}_{QGfE}$ given by 
\bea\hat{\mathcal{S}}_{QGfE}&=&\int d^4x \sqrt{-|\hat{g}|}\hat{\mathcal{L}}_{QGfE},
\eea
associated with the Lagrangian $\hat{\mathcal{L}}_{QGfE}$ that can be expressed as 
\bea
\hat{\mathcal{L}}_{QGfE}&=&\hat{\mathcal{L}}_{RM\Phi}-\hat{\mathcal{L}}_{M}.
\label{eq:hQGfE}
\eea
The Lagrangian $\hat{\mathcal{L}}_{RM\Phi}$ is quadratic in the $\Phi$ field and gravitationally neutral, i.e.
\bea
\hat{\mathcal{L}}_{RM\Phi}&=&\hat{\mathcal{L}}_{R\Phi}+\hat{\mathcal{L}}_{M\Phi}=\bar\Phi \tilde{\mathcal{K}}{\Phi}
\eea
with the kernel $\tilde{\mathcal{K}}$ being a function  of the matter fields $\hat{\Psi}$, the curvature $\tilde{\mathcal{R}}$ and the metric $\g$, i.e.
\bea
\tilde{\mathcal{K}}=\tilde{\mathcal{K}}[\hat{\Psi},\tilde{\mathcal{R}},\hat{g}].
\eea
Specifically, the kernel $\tilde{\mathcal{K}}$ can be expressed as
\bea
\tilde{\mathcal{K}}_{\mu\nu}[\hat{\Psi},\tilde{\mathcal{R}},\hat{g}]=\tilde{\square}\hat{g}_{\mu\nu}-M^2\tilde{\bf G}_{\mu\nu},
\label{eq:tK}\eea
where   $\tilde{\bf G}$ is the metric induced by the matter fields and curvature and is given by 
\bea
\tilde{\bf G}=\tilde{g}+\alpha \tilde{\bf M}-\beta\tilde{\bm{\mathcal{R}}},
\eea
with $\alpha$ and $\beta$  given by Eq.(\ref{eq:ab}).

\section{Emergent Gravity from Entropy}
\label{sec:EGfE}
In this section, the EGfE action is derived as the effective action of QGfE. To this end, we start from  the neutrally charged quadratic approximate expression of the QGfE Lagrangian $\hat{\mathcal{L}}_{QGfE}$ given by Eq.(\ref{eq:hQGfE}) valid for small $\Phi$ fields, and we expand the corresponding action $\hat{\mathcal{S}}_{QGfE}$ around the classical values of the metric and the matter fields. Specifically, we take 
\bea
\hat{g}_{\mu\nu}&=&g_{\mu\nu}+\varkappa\hat{h}_{\mu\nu},\nonumber\\
\hat{\Psi}&=&\Psi+\hat{\Xi},
\label{eq:hX}
\eea
with $\varkappa=a_{\varkappa}/{M},$ where $a_{\varkappa}$ is a dimensionless constant that for later convenience we take to be $a_{\varkappa}=\sqrt{8\pi/(n_c\alpha)}$. In order to illustrate the derivation, first,  all quantum fluctuations will be neglected,  obtaining in this way  the classical EGfE action $\mathcal{S}_{EGfE}^{(c)}$. Secondly,  the full EGfE action will be derived by including also the contributions coming from the quantum fluctuations of the metric and the matter fields.
\subsection{The classical  EGfE }
\subsubsection{Derivation of the classical EGfE }
The effective theory obtained from the classical metric and matter fields is the classical EGfE shown  here to be equal to the GfE action  minus the matter field action $\hat{\mathcal{S}}_M$.
In order to discuss this derivation, let us now consider the action $\hat{\mathcal{S}}_{QGfE}$ and replace the quantum metric $\hat{g}$ and the quantum matter fields $\hat{\Psi}$ with their classical values, i.e. let us put 
\bea
\hat{g}_{\mu\nu}=g_{\mu\nu},\quad \hat{\Psi}=\Psi.
\eea
We denote with $\mathcal{R}$ the higher-order curvature  $\tilde{\mathcal{R}}$ evaluated on the classical background while in this section we retain the notation $\tilde{g}$ for the higher-order metric defined in Eq.(\ref{eq:gt}) calculated over the classical background.
In this classical limit, $\hat{\mathcal{L}}_{RM\Phi}$ is given by \bea\hat{\mathcal{L}}_{RM\Phi}=\bar\Phi \tilde{\mathcal{K}}_c{\Phi},
\eea
where $\tilde{\mathcal{K}}_c$ is evaluated on the classical fields, i.e. $\tilde{\mathcal{K}}_c=\tilde{\mathcal{K}}[\Psi,{\mathcal{R}},g]$.
The classical effective action ${\mathcal{S}_{EGfE}^{(c)}}$ is defined as 
\bea
\frac{Z_{\Phi}[\Psi,{\mathcal{R}},g]}{Z_{\Phi}[0,0,g]}=e^{\mathcal{S}_{EGfE}^{(c)}},
\eea
where the partition function $Z_{\Phi}[\Psi,{\mathcal{R}},g]$ is defined as in Eq.(\ref{eq:Zh}) in terms of the action $\hat{\mathcal{S}}_{QGfE}$ with associated Lagrangian given by Eq.(\ref{eq:hQGfE}).
This definition of the EGfE action fully expresses the interpretation of GfE as deriving from an effective field theory, and reinterprets the role of the GQRE  between the metric induced by the matter fields and curvature and the true metric, as emerging from the above normalization of the partition function.
Both the partition function $Z_{\Phi}[\Psi,{\mathcal{R}},g]$ and the partition function $Z_{\Phi}[0,0,g]$ contain a Gaussian integral of the $\Phi$ field with kernel $i\tilde{\mathcal{K}}[\Psi,{\mathcal{R}},g]$ and kernel $i\tilde{\mathcal{K}}[0,0,g]$ respectively, where $\tilde{\mathcal{K}}[\Psi,{\mathcal{R}},g]$ is given by Eq.(\ref{eq:tK}).

Thus, performing the complex Gaussian integral and retaining only the leading term for $M\gg 1$ under the assumption that the $\Phi$ field has a momentum cutoff $\epsilon M$, with $\epsilon\ll 1$,  we obtain  the classical EGfE action, 
\bea
\mathcal{S}_{EGfE}^{(c)}=\int d^4x \sqrt{-|g|}\mathcal{L}_{EGfE}^{(c)},
\eea
associated with the classical EGfE Lagrangian $\mathcal{L}_{EGfE}^{(c)}$ given by
\bea
 \mathcal{L}_{EGfE}^{(c)}=\mathcal{L}_{GfE}-{\mathcal{L}}_M.
 \eea
In this expression $\mathcal{L}_{GfE}$ is given by the integral performed on the momentum ${\bf p}$ defined in the local frame:
\bea
 \mathcal{L}_{GfE}=-\int_B \frac{d^4p}{(2\pi)^4}\mbox{Tr}_F\ln\left(\frac{\tilde{\bf G}\tilde{g}^{-1}+p^2/M^2\tilde{\bf P}_{\pi}}{\tilde{\bf I}+p^2/M^2\tilde{\bf P}_{\pi}}\right)
\eea
where $\tilde{\bf P}_{\pi}$ is the projector on the $\pi$ sector of the $\Phi$ field, and the integral is performed on the region \bea
B=\left\{{\bf p} \Big| |p_0|<\epsilon M \ \mbox{and}\  {|\vec{p}|<\epsilon M}\right\}.
\eea
 We thus obtain, to leading order in the large-$M$ expansion, the expression of the GfE Lagrangian  proposed in \cite{bianconi2025gravity} given by  the GQRE between the true metric $\tilde{g}$ and the metric induced by the matter fields and curvature given by $\tilde{\bf G}$, i.e. 
\bea
\mathcal{L}_{GfE}=-n_c\mbox{Tr}_F\ln (\tilde{\bf G}\tilde{g}^{-1}).
\eea
Moreover, the matter Lagrangian is given by
\bea
\mathcal{L}_M=-\frac{n_c\alpha}{2}\mbox{Tr}_F\tilde{\bf M}\tilde{g}^{-1},
\eea
where it is to be noted that $n_c\alpha=O(M^0)$.
Interestingly, we can investigate the classical EGfE for low momenta of the matter fields and for small curvature, by linearizing the logarithm in the GfE Lagrangian. In particular, by assuming 
 \bea
{\alpha^{\prime}}=4\pi\Big(4\beta^{\prime}(a_0+a_1+1)-1\Big)>0,
\eea we obtain  that in the weak-field limit, the classical EGfE Lagrangian can be approximated  by the GR Lagrangian $\mathcal{L}_{GR}$, i.e.
\bea
\mathcal{L}_{EGfE}^{(c)}\simeq a_{GR}\mathcal{L}_{GR}\eea
with
\bea
\mathcal{L}_{GR}&=&\frac{M^2}{16\pi}R-m_\chi^2|\chi|^2-|\bar\nabla\chi|^2-\frac{1}{4}F_{\mu\nu}F^{\mu\nu}.
\eea
Under this assumption, the proportionality constant $a_{GR}=O(M^0)$ is given by $a_{GR}=4n_c\alpha={2\alpha^{\prime}}\epsilon^4/({3\pi^3}).$
\subsubsection{Formulation of the classical EGfE in terms of   the G-field}

The GfE action depends exclusively on $\tilde{\bf G}\tilde{g}^{-1}$. Thus, following the treatment proposed in Ref.\cite{bianconi2025gravity}, we introduce  an auxiliary field $\tilde{\bm{\aleph}}$ and  we impose the constraints
\bea
\tilde{\bf G}\tilde{g}^{-1}=\tilde{\bm{\aleph}}.
\label{constraints}
\eea
 with Lagrange multipliers $\tilde{\bm{\mathcal{G}}}$ defining the G-field, thereby reducing the classical EGfE to 
\bea
{\mathcal{L}}^{(c)}_{EGfE}=n_c\mbox{Tr}_F\left[-\ln \tilde{\bm{\aleph}}-\tilde{\bm{\mathcal{G}}}(\tilde{\bf G}\tilde{g}^{-1}-\tilde{\bm\aleph})+\frac{\alpha}{2}\tilde{\bf M}\tilde{g}^{-1}\right].\nonumber 
\label{LW}
\eea
By imposing that the variation with respect to $\tilde{\bm{\aleph}}$ vanishes, we get 
\bea
\tilde{\bm{\aleph}}^{-1}=\tilde{\bm{\mathcal{G}}}.\label{inv}
\eea
Thus, this relation implies
\bea
\tilde{\bm{\mathcal{G}}}^{-1}=\tilde{\bf I}+\alpha\tilde{\bf M}\tilde{g}^{-1}-\beta\tilde{\bm{\mathcal{R}}}\tilde{g}^{-1},
\label{Gm1b}
\eea
and we can simply eliminate the field $\tilde{\bm\aleph}$  from the action.
In this way  the considered Lagrangian ${\mathcal{L}}_{EGfE}^{(c)}$ can be written as
\bea
{\mathcal{L}}_{EGfE}^{(c)}=n_c\alpha\left[\frac{\beta}{\alpha}\Big(\mathcal{R}_{\mathcal{G}}-2\Lambda_{\mathcal{G}}\Big)-{\mathcal{M}}_{\mathcal{G}}\right],
\label{St}
\eea

where $\mathcal{R}_{\mathcal{G}}$ is the {\em dressed Ricci scalar} and $\mathcal{M}_{\mathcal{G}}$ includes all dependence on the matter fields while $\Lambda_{\mathcal{G}}$ is the emergent  positive dark energy term. Specifically, defining the {\em dressed metric} $\tilde{g}_{\mathcal{G}}$ as $\tilde{g}_{\mathcal{G}}=\tilde{\bm{\mathcal{G}}}^{-1}\tilde{g}$, we have
\bea
\mathcal{R}_{\mathcal{G}}=\mbox{Tr}_{F}\tilde{g}_{\mathcal{G}}^{-1}\tilde{\bm{\mathcal{R}}},\quad 
\mathcal{M}_{\mathcal{G}}= \mbox{Tr}_{F}\left[\left(\tilde{g}_{\mathcal{G}}^{-1}-\frac{1}{2}{\tilde{g}}^{-1}\right) \tilde{\bf M}\right],\nonumber\eea
while the emergent dark energy term is given by 
\bea
\Lambda_{\mathcal{G}}=\frac{1}{2\beta}\mbox{Tr}_{F}\Big(\tilde{\bm{\mathcal{G}}}-\tilde{\bf I}-\ln\tilde{\bm{\mathcal{G}}}\Big).
\label{dressed_eq}
\eea

\subsubsection{Classical metric}
The classical  values  of the metric and the matter fields are chosen to be the solutions of the variational equations 
\bea
\left.\frac{\delta \mathcal{S}^{(c)}_{EGfE}}{\delta g_{\mu\nu}}\right|_{\tilde{\bm{\mathcal{G}}}=\tilde{\bm{\mathcal{G}}}_c}=0,\quad \left.\frac{\delta \mathcal{S}_{EGfE}^{(c)}}{\delta \Psi}\right|_{\tilde{\bm{\mathcal{G}}}=\tilde{\bm{\mathcal{G}}}_c}=0,
\eea
where variations with respect to conjugate matter fields are implicit.
Consequently, it  follows that  the  modified gravity equations derived from the classical EGfE differ from the original GfE equations of motion \cite{bianconi2025gravity} only by a redefinition of the stress-energy tensor.

\subsection{Including quantum contributions in EGfE}\subsubsection{The Euclidean partition function}
In this section we illustrate the effect of taking into account the  small quantum fluctuations of the metric and the matter fields. We define  the EGfE action as 
\bea
e^{-\mathcal{S}_{EGfE}^E}=\int \mathcal{D}\hat{h}\int \mathcal{D}\bar{\hat\Xi}\int \mathcal{D}\hat{\Xi}\frac{Z^E_{\Phi}[\hat\Psi,\tilde{\mathcal{R}},\g]}{Z^E_{\Phi}[0,0,\g]}
\label{eq:EGfEq}
\eea
where we assume that the measure over $\hat{h}$ is defined in such a way to correctly account for diffeomorphism invariance.
The gauge fixing associated with the internal gauge symmetries of the matter fields is also understood in the definition of the functional integral.
In this expression we have introduced the Euclidean partition function ${Z}^E_{\Phi}[\hat\Psi,\tilde{\mathcal{R}},\g]$ obtained from  $Z_{\Phi}[\hat\Psi,\tilde{\mathcal{R}},\g]$ by performing a Wick rotation:
 \bea
{Z}^E_{\Phi}[\hat\Psi,\tilde{\mathcal{R}},\g]=\int \mathcal{D}\bar\Phi\int \mathcal{D}\Phi  e^{-\hat{\mathcal{S}}_{QGfE}^E},\label{eq:ZEPhi}
\eea

The Euclidean action $\hat{\mathcal{S}}_{QGfE}^E$ is associated with the Euclidean Lagrangian $\hat{\mathcal{L}}_{QGfE}^E$  by 
\bea
\hat{\mathcal{S}}_{QGfE}^E=\int d^4x\sqrt{|\g|}\hat{\mathcal{L}}^E_{QGfE},
\eea
where $\hat{\mathcal{L}}_{QGfE}^E$ is obtained from Eq.(\ref{eq:hQGfE}) by performing a Wick rotation and is given by
\bea
\hat{\mathcal{L}}_{QGfE}^E=\hat{\mathcal{L}}_{RM\Phi}^E-\hat{\mathcal{L}}^E_M.
\label{eq:hQGfE2}
\eea
The Lagrangian $\hat{\mathcal{L}}_{RM\Phi}^E$ can thus be expressed as \bea\hat{\mathcal{L}}_{RM\Phi}^E=\bar\Phi \tilde{\mathcal{K}}^E{\Phi}=\bar\Phi(-\tilde{\square}_E \g+M^2\tilde{\bf G}_E)\Phi,\label{eq:tK2}
\eea
where $\tilde{\square}_E$ is obtained from $\tilde{\square}$ by performing the Wick rotation and thus has non-positive eigenvalues.

\subsubsection{Quantum corrections to classical EGfE}
Taking into account the quantum fluctuations of the metric and the matter fields we obtain the quantum Euclidean EGfE action $\mathcal{S}_{EGfE}^{(q,E)}$ defined as
\bea
{\mathcal{S}_{EGfE}^{(q,E)}}=-\ln \left(\frac{Z^E_{\Phi}[\hat\Psi,\tilde{\mathcal{R}},\g]}{Z^E_{\Phi}[0,0,\g]}\right).
\eea
The action $\mathcal{S}_{EGfE}^{(q,E)}$ is naturally associated with the Lagrangian $\mathcal{L}_{EGfE}^{(q,E)}$, by 
\bea
\mathcal{S}_{EGfE}^{(q,E)}=\int d^4x \sqrt{|\g|}\mathcal{L}_{EGfE}^{(q,E)},\eea
where  $\mathcal{L}_{EGfE}^{(q,E)}$  in the large-$M$ limit is given by
\bea
\hspace{-5mm}\mathcal{L}_{EGfE}^{(q,E)}=n_c\mbox{Tr}_F\ln(\tilde{\bf G}_E\tilde{g}^{-1}+\delta\tilde{\bf G}_E\tilde{g}^{-1})-\hat{\mathcal{L}}_{M}^E[\hat\Psi,\g],
\eea
where here (see Appendix \ref{ApB} for details) we identify the number of field modes per unit proper volume $n_c^E$ obtained in this Euclidean field theory with $n_c$.
Here $\delta \tilde{\bf G}_E$ indicates the change in $\tilde{\bf G}_E$ due to the quantum fluctuations of the metric and the matter fields, i.e.
\bea
\delta \tilde{\bf G}_E=\tilde{\bf G}_E[\hat\Psi,\tilde{\mathcal{R}},\g]-\tilde{\bf G}_E[\Psi,{\mathcal{R}},g].
\eea
Using the definition of the EGfE action given by Eq.(\ref{eq:EGfEq}) we thus obtain  
\bea
\mathcal{S}_{EGfE}^E=\mathcal{S}_{EGfE}^{(c,E)}+\Gamma_{EGfE}^E.
\eea
Here  $\Gamma_{EGfE}^E$  denotes the effective action  associated with the integration of the quantum fluctuations of the metric and the matter fields. In order to express $\Gamma_{EGfE}^E$ we follow standard GR perturbative calculations \cite{t1974one,kiefer,buchbinder2021introduction}. Specifically, we   introduce  a gauge-fixing Lagrangian $\mathcal{L}_{gauge}^E$ and  the Faddeev-Popov  complex-valued ghost $c$ with its associated Lagrangian $\mathcal{L}_{ghost}^E$, thus leading to 
\bea
{e^{-\Gamma_{EGfE}^E}} &=&\int \mathcal{D}\hat\Xi\int \mathcal{D}\bar{\hat\Xi}\int \mathcal{D}\hat{h}\int \mathcal{D}c\int \mathcal{D}\bar{c} e^{-{\mathcal{S}}_{Q}^E},\nonumber 
\label{eq:ZGR}
\eea 
with $\mathcal{S}_{Q}^E$ given by
\bea
{\mathcal{S}}_{Q}^E={\mathcal{S}}_{EGfE}^{(q,E)}-{\mathcal{S}}_{EGfE}^{(c,E)}+\mathcal{S}_{gauge}^E+\mathcal{S}_{ghost}^E.
\eea
Here, adopting the covariant de Donder gauge-fixing function, we have  introduced \bea
\mathcal{L}_{gauge}^E&=&\frac{1}{4}F_{\mu}F^{\mu},\quad F_{\mu}=\nabla^{\nu}\hat{h}_{\mu\nu}-\frac{1}{2}\nabla_{\mu}\hat{h}_{\nu}^{\nu},\nonumber \\
\mathcal{L}_{ghost}^E&=&\bar{c}^{\mu}(-\square_E\delta_{\mu}{}^{\nu}-R_{\mu}{}^{\nu})c_{\nu}.
\eea
It follows that this Euclidean formulation inherits the well-known conformal-factor problem of Euclidean general relativity. At the perturbative level, the  standard conformal contour prescription introduced in Ref. \cite{gibbons1978path} should therefore be adopted, whereby the integration contour of the negative conformal modes is rotated into the complex plane. Beyond the weak-field limit, we restrict our perturbative treatment to classical backgrounds for which the gauge-fixed quadratic EGfE action admits a well-defined Gaussian integral on an appropriate integration contour.

We note that in the weak-field limit the G-field $\tilde{\bm{\mathcal{G}}}$ can be approximated by the higher-order identity. In this limit, to leading order in the low-momentum expansion, $\Gamma_{\mathrm{EGfE}}^E$, evaluated  at quadratic order in the metric and matter-field fluctuations, reduces to the one-loop effective action of Euclidean  GR coupled to matter \cite{t1974one}. Importantly, since EGfE is derived as a low-energy effective theory of QGfE, the integration over the quantum fluctuations of the metric and the matter fields is naturally restricted to energy scales much smaller than $M$. Consistently with the derivation of classical EGfE, we adopt an ultraviolet cutoff $\epsilon_E M$ with $\epsilon_E\ll 1$ in the calculation of $\Gamma_{{EGfE}}^E$. This cutoff is physically motivated by the regime of validity of the effective theory and is not intended to be removed.  Within this framework, the fundamental question of renormalizability concerns the underlying QGfE theory rather than EGfE or its weak-field limit given by GR.

\subsubsection{Lorentzian EGfE}
Having derived the Euclidean EGfE effective action, we can recover the Lorentzian formulation of EGfE by performing the inverse Wick rotation, obtaining
\bea
\mathcal S_{EGfE}=\mathcal S_{EGfE}^{(c)}+\mathcal{S}_{QC}\eea
In particular, the classical contribution reduces to the Lorentzian classical EGfE action $\mathcal{S}_{EGfE}^{(c)}$, while the quantum corrections $\mathcal{S}_{QC}$ are obtained by analytically continuing the corresponding Euclidean effective action, i.e.
\bea
\mathcal{S}_{QC}=i\Gamma_{EGfE}^E\big|_{\tau=it}.
\eea
While the continuation is straightforward for the local contributions to the effective action, nonlocal contributions require the specification of the appropriate analytic continuation and boundary conditions.

\section{Conclusions}
\label{sec:conclusions}

In this work we have proposed QGfE, a quantum gravity theory driven by the massive GF that in the fundamental state reproduces or shadows the true metric, and whose deviations from the fundamental state, captured by the $\Phi$ field, couple to both matter and curvature. Interestingly, the QGfE does not contain any Einstein-Hilbert term, or in fact, any free terms for either the matter fields or the curvature, which only appear in the action through their coupling with the $\Phi$ field.
Close to the fundamental state, the $\Phi$ field  describes the quantum fluctuations of the GF metric, which are  massive, with a mass given by the Planck mass $M=M_P$. Note, however, that a massive $\Phi$ field in QGfE remains totally compatible with the existence of the massless graviton obtained in the weak field limit of its effective action.
We  explore the QGfE action when the GF is close to its fundamental state and $\Phi$ is small, and its energy scales are below $\epsilon M$ with $\epsilon\ll 1$. By also considering an expansion of the true metric and the matter fields around their classical values,  we derive the QGfE effective action which we call the EGfE action.
When the fluctuations of the metric and the matter fields are neglected, we obtain the classical EGfE, given by the GfE action minus the matter action. Around the classical solution, the variation of the EGfE action  emerges as the action which dictates the quantum fluctuations of the metric and the matter fields.
Integrating these quantum fluctuations leads to the full EGfE action, given by the classical EGfE action supplemented by the effective action associated with the quantum fluctuations of the metric and the matter fields. A further important direction is the extension of the proposed framework to fermionic matter and ultimately to the full Standard Model.

This work opens different perspectives.
Firstly, the GF emerges as a promising candidate to explain the dark sector. 
Indeed, the $\Phi$ field is an additional massive field that might contribute to dark matter, while the G-field $\tilde{\bm{\mathcal{G}}}$, which modulates the fluctuations of the $\Phi$ field, is responsible for the GfE dark energy term $\Lambda_\mathcal{G}$. Secondly, by deriving EGfE as the effective field theory of QGfE, this work proposes QGfE as a candidate high-energy completion of GfE and thus, in the weak-field limit, of GR itself. Thirdly, EGfE  emerges as the action driving the quantum perturbations around the classical EGfE, thus providing a mechanism according to which in the weak-field limit, GR itself is emergent. The interpretation of EGfE and, in the weak-field limit, GR as effective field theories shifts the fundamental question of renormalizability to the underlying QGfE theory. Finally, given that the QGfE Lagrangian contains only dimensionless coupling constants, the investigation of its renormalization group properties and ultraviolet behavior represents a natural direction for future research.

\section{Acknowledgments}
This publication was made possible through the support of grant \#63793 from the John Templeton Foundation.
This research benefited from discussions with L. Freidel, D. Litim, R. Percacci and C. Wetterich at the conferences ``Concepts of Quantum and Spacetime" Tsukuba, Japan (2026), ``Exact Renormalization Group" (ERG2026), Brighton, UK and ``Spacetime-Matter-Quantum Mechanics" (DICE2026), Castiglioncello, Italy.
\bibliography{references.bib}
\appendix

\section{Main steps in the derivation of $\hat{\mathcal{S}}_{M\Phi}$}
\label{ApA}
In this Appendix, further details are provided on the derivation of the leading neutrally charged contribution to the quadratic expansion of the QGfE action close to its fundamental state. We start from the bilocal action ${\mathcal{S}}_{M\Phi}^{\prime}$, which we recall here for convenience:
\bea
{\mathcal{S}}_{M\Phi}^{\prime}=\int d^4x \sqrt{-|\hat{g}|}\int d^4y \sqrt{-|\hat{g}|}\mathcal{F}_B(x,y).
\eea
Interestingly, $\mathcal{F}_B(x,y)$ can be expressed as  
\bea
\mathcal{F}_B(x,y)&=&-{i}{\alpha^{\prime}}\hat{Q}_{NN'}(x,y)\bar{J}^{N'}(x){J}^N(y)
\eea
where $\hat{Q}_{NN'}$ is fourth order in the $\Phi$ field and is given by
\bea
\hat{Q}_{NN'}=\bar{\pi}_{\mu}(x)\bar\Phi^{\mu}_{N'}(x)\pi_{\nu}(y)\Phi^{\nu}_N(y).
\eea
By performing a Hartree approximation on $\hat{Q}_{NN'}$, taking into account that, in our notation $\pi_{\mu}=\Phi_{\mu}^0$, we obtain 
\bea
\hat{Q}_{N,N'}(x,y)&\simeq&\Delta_{\mu\nu}^{00}(x-y)\bar\Phi_{N'}^{\mu}(x)\Phi_N^{\nu}(y)\nonumber\\
&&+\;\Delta_{NN'}^{\mu\nu}(x-y)\;\bar\pi_{\mu}(x)\pi_{\nu}(y)\nonumber \\&&+\delta_{0N}\Delta^{00}_{\mu\nu}(x-y)\bar\Phi_{N'}^{\mu}(x)\pi^{\nu}(y) \nonumber\\
&&+\;\delta_{0N'}\Delta_{\mu\nu}^{00}(x-y)\;\bar\pi^{\mu}(x)\Phi_{N}^{\nu}(y)\nonumber \\&&-\mathcal{C}_{N,N'}(x-y)
\eea
where 
\bea
\mathcal{C}_{N,N'}(x-y)&=&\left[\Delta_{\mu\nu}^{00}(x-y)\Delta^{\mu\nu}_{NN'}(x-y)\right.\nonumber \\
&&\left.+\delta_{0N}\delta_{0N^{\prime}}\Delta_{\mu\nu}^{00}(x-y)\Delta^{\mu\nu}_{00}(x-y)\right].\nonumber
\eea
Using the expression for the propagators of the $\Phi$ field, given by Eq.(\ref{eq:Dprop}), we can approximate  ${\mathcal S}^{\prime}_{M\Phi}$ with a local quadratic term in $\Phi$ plus a constant term in $\Phi$, i.e.
\bea
\hat{\mathcal S}_{M\Phi}=\int\sqrt{-|\g|}d^4x\left[\bar\Phi_{N'}^{\mu}(x)\,\tilde{\mathcal {K}}^{N'N}_{\mu\nu}(x)\,\Phi_N^{\nu}(x)-{i}\hat{\mathcal{L}}_B(x)\right],\nonumber
\eea
with the local kernel
\bea
\tilde{\mathcal{K}}^{N'N}_{\mu\nu}(x)&=&-\frac{\alpha^{\prime}}{M^2}\hat{g}_{\mu\nu}\left[\bar{J}^{N'}(x)J^N(x)+\delta^{N}_{0}\bar{J}^{N'}(x)J^0(x)\right.\nonumber\\&&\left.\hspace{-10mm}+\delta_{0}^{N'}\bar{J}^{0}(x)J^{N}(x)+\delta^{N'}_{0}\delta^{N}_{0}\bar{J}_M(x)J^{M}(x)\right],
\label{eq:kernel2}
\eea
where $\hat{\mathcal{L}}_{B}$ is given by
\bea
\hat{\mathcal{L}}_{B}(x)=\frac{4n_c\alpha^{\prime}}{M^4}\left[\bar{J}^{N}(x)J_N(x)+\bar{J}^{0}(x)J_{0}(x)\right].
\eea
where $n_c$ indicates the number of field modes per unit proper volume (see Appendix \ref{ApB} for the derivation).

\section{Number of field modes per unit proper volume}
\label{ApB}
\subsubsection{Lorentzian case}
In Lorentzian spacetime, the number of field modes $n_c$ at coincident points, is calculated by introducing a cutoff $\epsilon M$ with $\epsilon\ll 1$. Specifically, it can be expressed in terms of the components ${\bf p}=(p_0,\vec{p})$ of the momentum in the local frame at $x$, by integrating the density of states associated with these modes on the region $B$ given by 
\bea
B=\left\{{\bf p} \Big| |p_0|<\epsilon M \ \mbox{and}\  {|\vec{p}|<\epsilon M}\right\}.
\eea
Thus,  we obtain
\bea
n_c=\int_{B} \frac{d^4p}{(2\pi)^4}=\frac{\epsilon^4M^4}{6\pi^3}.
\eea
\subsubsection{Euclidean case}
In Euclidean spacetime, obtained after performing a Wick rotation, the number of field modes $n_c^E$ at coincident points, is calculated by introducing a cutoff $\epsilon_E M$ with $\epsilon_E\ll 1$. Specifically, it can be expressed in terms of the components $p_{\mu}$ of the momentum in the local frame at $x$, by integrating the density of states associated with these modes on the region $B_E$ given by 
\bea
B_E=\left\{{\bf p} \Big|   {|{\bf p}|<\epsilon_E M}\right\}.
\eea
Thus,  we obtain
\bea
n_c^E=\int_{B_E} \frac{d^4p}{(2\pi)^4}=\frac{\epsilon_E^4M^4}{32\pi^2}.
\eea
Here, for simplicity, we assume that $\epsilon_E^4=16\epsilon^4/(3\pi)$, and  
thus 
\bea
n_c^E=n_c.\eea
\end{document}